\documentclass[11pt]{article}
\usepackage{moriond,epsfig}
\usepackage{amsmath}
\usepackage{subfigure}
\def\MyPhoto{\psfig{figure=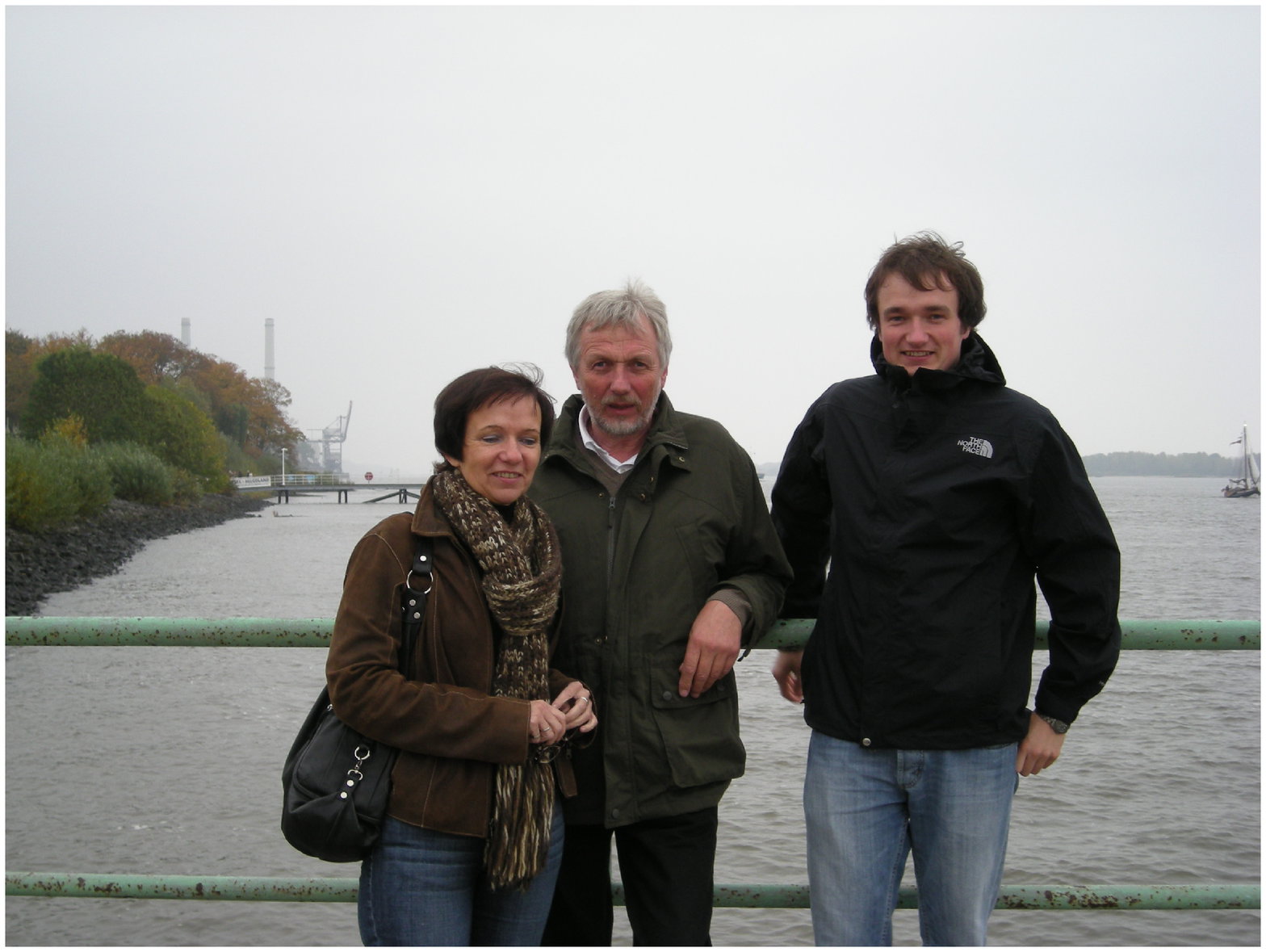,bb= 385 252 470 345, clip=, height=31mm}}

\renewcommand{\thesubfigure}{\arabic{subfigure}}
\makeatletter
   \renewcommand{\p@subfigure}{}
   \renewcommand{\@thesubfigure}{Fig. \thesubfigure:\hskip\subfiglabelskip}
\makeatother

\def\Journal#1#2#3#4{{#1} {\bf #2}, #3 (#4)}

\def\NPB{{\em Nucl. Phys.} B}
\def\PLB{{\em Phys. Lett.}  B}
\def\PRL{\em Phys. Rev. Lett.}
\def\PRD{{\em Phys. Rev.} D}

\def\be{\begin{equation}}
\def\ee{\end{equation}}
\def\bea{\begin{eqnarray}}
\def\eea{\end{eqnarray}}

\begin{document}
\vspace*{4cm}
\title{Global Fits of the electroweak Standard Model and beyond with Gfitter}

\author{ Martin Goebel~\footnote{on behalf of the Gfitter Group (www.cern.ch/Gfitter)}}

\address{DESY and Institut f\"ur Experimentalphysik der Universit\"at Hamburg,\\
Notkestra{\ss}e 85, 22607 Hamburg, Germany }

\maketitle
\abstracts{In the global fit of the SM using Gfitter, electroweak precision observables as well as 
constraints from direct Higgs searches have been compared with state-of-the-art electroweak predictions.
We use the most recent results for direct Higgs searches at LEP and Tevatron and the latest measurements
of $m_t$ and $M_W$. Example results are an estimation of the mass of the Higgs boson 
($M_H\ =\ 116.3^{\,+15.6}_{\,-1.3}\,{\rm GeV}$) and a forth-order result for the strong coupling constant 
($\alpha_S(M_Z^2)=0.1193 \pm 0.0028 {\rm(exp)} \pm 0.0001{\rm (theo)}$).
A fit of the oblique parameters ($STU$) to the electroweak data is performed, in order to constrain physics beyond the 
Standard Model. For instance, the parameter space of the Littlest Higgs Model with T-parity can be restricted via 
the oblique parameters. In addition, fit results for a model with an extended Higgs sector (2HDM) 
using mainly observables from the $B$ and $K$ physics are presented.}

\section{Introduction}
Precision measurements allow us to probe physics at much higher energy scales than 
the masses of the particles directly involved in experimental reactions by 
exploiting contributions from quantum loops. Prominent examples are the electroweak 
precision measurements, which are used in conjunction with the Standard Model (SM)
to predict via multidimensional parameter fits unmeasured quantities like the Higgs mass.
%In addition, the small deviations between the data and the theoretical predictions 
%of the SM set constraints in the parameter space of new physics models. A common approach 
%to describe those deviations are to parametrize virtual loop corrections
%from new physics models by the oblique parameters. \par

Several theoretical libraries within and 
beyond the SM have been developed in the past containing the pertubative calculations of the SM and new 
physics models for the electroweak observables. However, most of these programs are 
relatively old, were implemented in outdated programming languages, and are difficult 
to maintain with respect to the theoretical and experimental progress expected during the
forthcoming era of the LHC. These considerations led to development of the generic 
fitting package {\it Gfitter}~\cite{gfitter}, designed to provide a modular framework for complex fitting 
tasks in high-energy physics. {\it Gfitter} is implemented in C++ and relies on 
ROOT functionality. The package allows a consistent treatment of statistical, 
systematic and theoretical errors, possible correlations and inter-parameter
dependencies. \par

In this paper we present {\it Gfitter} results of the global SM fit to the electroweak observables
as well as an estimate of the oblique parameters, which can be also used to constrain the 
parameter space of the Littlest Higgs Model (LHM). In addition, a fit of a 
Two Higgs Doublets Model (2HDM) is performed using {\it B} and {\it K} physics observables.    

% SM fit
\section{The Global SM Fit\label{smfit}}
In the global electroweak fit the state-of-the-art calculations of the electroweak precision 
observables are compared with the most recent experimental data to constrain the free 
parameters of the fit and to test the goodness-of-fit. The free parameters of the SM relevant for 
the global electroweak analysis are the coupling constant of the electromagnetic, weak, 
and strong interactions, as well as the masses of the elementary fermions and bosons. 
Due to electroweak unification and simplifications arising from fixing parameters with 
insignificant uncertainties compared to the sensitivity of the fit, the number of free fit 
parameters can be reduced. The remaining floating parameters in the fit are the coupling parameters 
$\Delta\alpha^{(5)}_{\rm had}(M_Z^2)$ and $\alpha_S(M_Z^2)$, the masses $M_Z$,
$\overline{m}_c$, $\overline{m}_b$, $m_{\rm top}$, and $M_H$.\par

In {\it Gfitter} a complete new library of the electroweak precision observables as 
measured by the LEP, SLC, and Tevatron experiments has been implemented. 
State-of-the-art predictions in the one-mass-shell scheme are used. In particular,
the full two-loop and leading beyond two-loop corrections are available for the
predictions of $M_W$~\cite{awramik1} and $\sin^2\theta^l_{\rm eff}$~\cite{awramik2,awramik3}. 
The implementation of the NNNLO pertubative calculation of the massless QCD Adler 
function~\cite{adler}, contributing to the 
vector and axial radiator functions in the prediction of the $Z$ hadronic width, allows
to fit the strong coupling constant with a unique theoretical accuracy. Wherever possible 
the calculations have been cross-checked against the ZFITTER package~\cite{zfitter}. More details on 
the theoretical computations in {\it Gfitter} can be found in~\cite{gfitter}. \par 

The following experimental measurements are used: The mass and width of the $Z$ boson, 
the hadronic pole cross section $\sigma^0_{\rm had}$, the partial widths ratio $R^0_l$, 
and the forward-backward asymmetries for leptons $A_{\rm FB}^{0,l}$, have been determined 
by fits to the $Z$ line-shape measured precisely at LEP (see~\cite{zsummary} and references 
therein). Measurements of the $\tau$ polarization at LEP~\cite{zsummary} and the left-right 
asymmetry at SLC~\cite{zsummary} have been used to determine the lepton asymmetry parameter 
$A_l$. The corresponding $c$ and $b$-quark asymmetry parameters $A_{c(b)}$, the 
forward-backward asymmetries $A_{\rm FB}^{0,c(b)}$, and the widths ratios $R^0_c$ and $R^0_b$, 
have been measured at LEP and SLC~\cite{zsummary}. In addition, the forward-backward 
charge asymmetry measurement in inclusive hadronic events at LEP was used to directly 
determine $\sin^2\theta_{\rm eff}^{l}$~\cite{zsummary}. For the running quark masses 
$\overline{m}_c$ and $\overline{m}_b$ the world average values are used. For 
$\Delta\alpha_{\rm had}^{(5)}(M_Z^2)$ we take the phenomenological result~\cite{hagiwara}. 
For the $W$ width we use the official combined LEP and Tevatron result, while for the $W$
mass we take also into account the recent D0 measurement~\cite{newwmass} leading to our private preliminary 
combined value of $M_{W}= (80.399\pm0.023)$ GeV. In case of the top mass the latest combined 
result~\cite{newtopmass} $m_t=(173.1\pm1.3)$ GeV is used. 
The direct searches for the SM Higgs Boson at LEP~\cite{higgsLEP} and the most recent 
results from the Tevatron~\cite{higgstev}, leading to a 95\% confidence 
level (CL) exclusion for $M_H<114.4$~GeV and at $M_H=[160,170]$~GeV respectively, 
are included using a Gaussian approach that quantifies the difference between the 
observed test statistics (the log-likelihood ratios) and the expected values 
for the s+b hypothesis using the values of the respective confidence level 
(${\rm CL}_{\rm S+B}$). A contribution to the $\chi^2$ estimator of the fit is derived for 
each Higgs mass. More details of the procedure can be found in~\cite{gfitter}.
We perform global fits in two versions: the {\it standard (``blue-band'') fit} makes
use of all the available information except for the direct Higgs searches and
the {\it complete fit} uses also the constraints from the direct Higgs searches. \par
\begin{figure}
\centering
\subfigure[$\Delta \chi^2$ as a function of $M_H$ for the {\it complete fit}.]
   {\epsfig{figure=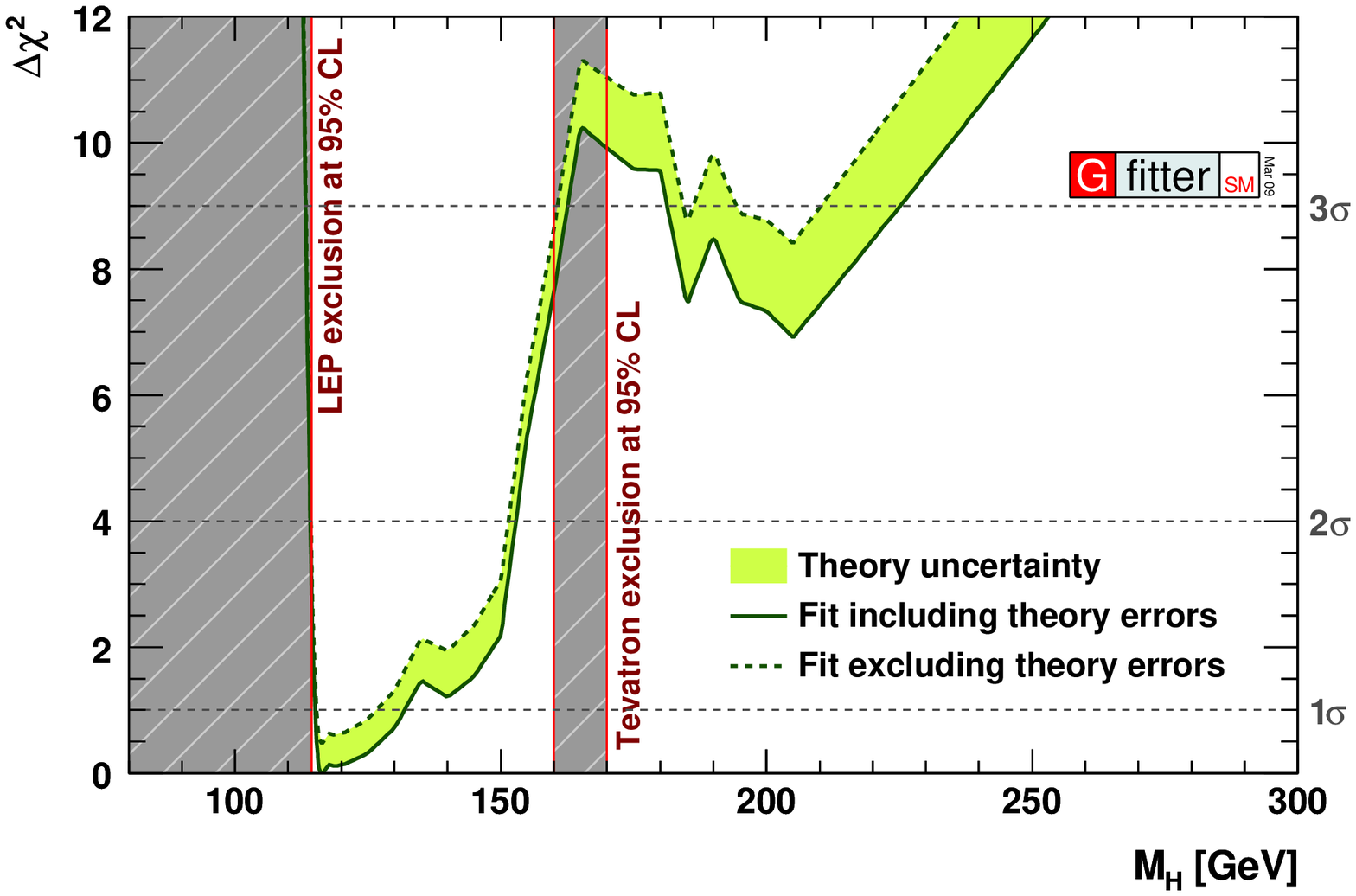, scale=0.38}\label{sm1}}\hfill
\subfigure[Determination of $M_H$ excluding all the sensitive observables from the {\it standard fit},
            except for the one given.]
   {\epsfig{figure=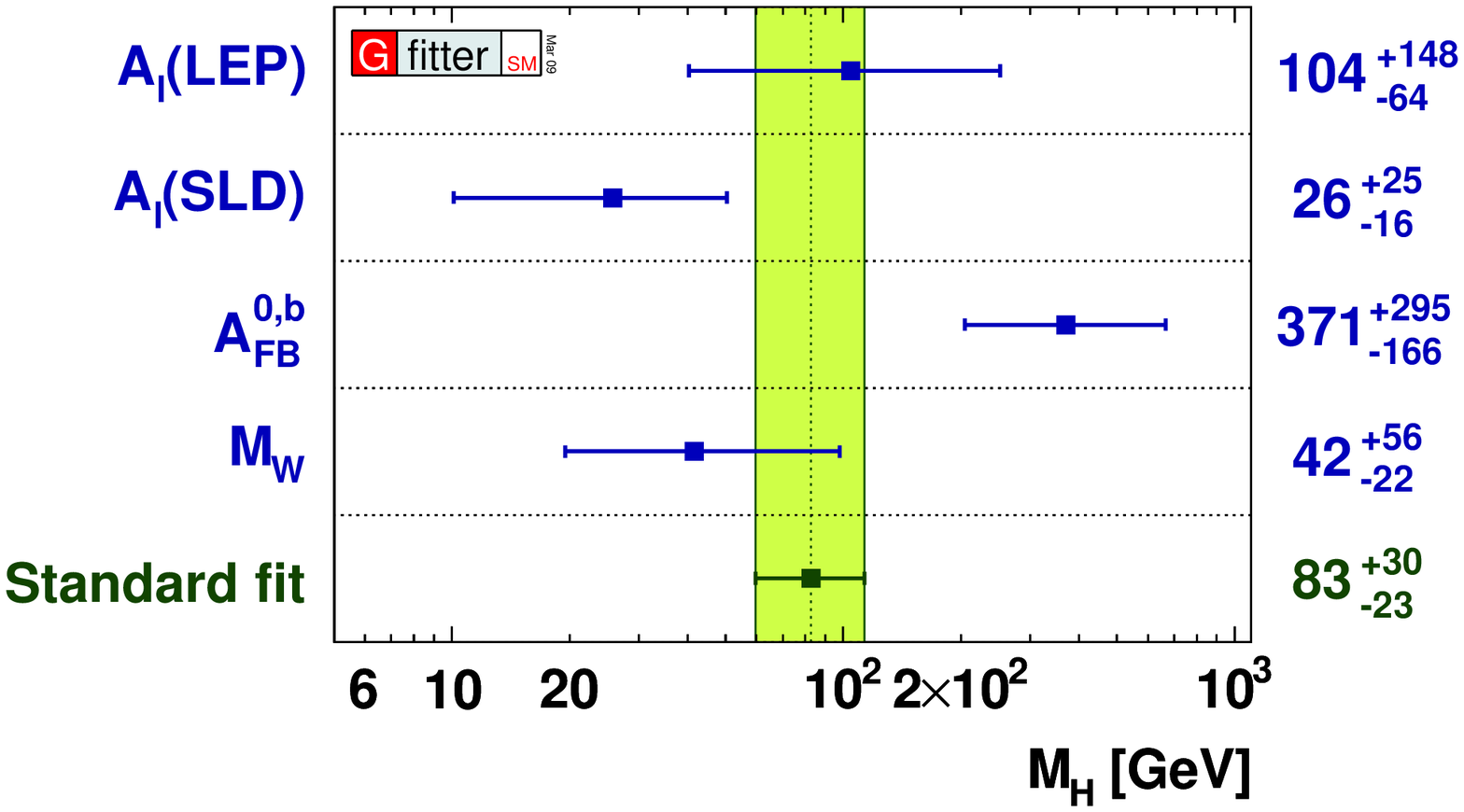,  scale=0.38}\label{sm2}}
\end{figure}
The {\it standard} ({\it complete}) {\it fit} converges at the global minimum 
value $\chi^2_{\rm min}=16.4$ ($\chi^2_{\rm min}=17.9$) for 13 (14) degrees of freedom, 
corresponding to a p-value of $0.228\pm0.004_{-0.002}$ ($0.204\pm0.004_{-0.002}$) derived from
MC toy experiments. The estimation for $M_H$ from the {\it standard fit}, {\it i.e.}, 
without the direct Higgs searches is $M_H\ =\ 83^{\,+30}_{\,-23}\:\rm GeV$ and the $2\sigma$ 
and $3\sigma$ intervals are respectively $[42,\,158]\,\rm GeV$ and $[28,\,211]\,\rm GeV$. 
The {\it complete fit} represents the most accurate estimation of $M_H$ considering 
all available data. The resulting $\Delta\chi^2$ curve versus $M_H$ is shown in Figure~\ref{sm1}.
The shaded band indicates the influence of theoretical uncertainties, 
which are included in the fit with a flat likelihood within the allowed ranges. 
The inclusion of the direct Higgs search results from LEP leads to a strong rise of the 
$\Delta\chi^2$ curve below $M_H=115\,\rm GeV$. The data points from the searches at the 
Tevatron, available in the range $110\, {\rm GeV} < M_H < 200\,{\rm GeV}$ increases the 
$\Delta\chi^2$ estimator for Higgs masses above $150\,{\rm GeV}$ beyond that obtained 
from the {\it standard fit}. The estimation for $M_H$ from the {\it complete fit} 
results in $M_H\ =\ 116.3^{\,+15.6}_{\,-1.3}$~GeV and the $2\sigma$ interval is reduced 
to $[114,\,145]\,\rm GeV$. \par

In figure~\ref{sm2} only the observable indicated in a given row of the plot is included in the fit.
Only for the four observables providing the strongest constraint on $M_H$, namely $A_\ell$(LEP), 
$A_\ell$(SLD), $A_{\rm FB}^{0,b}$ and $M_{W}$, the Higgs mass is determined. 
The compatibility among these measurements can be estimated
by repeating the global fit where the least compatible of the measurements (here
$A_{\rm FB}^{0,b}$) is removed, and by comparing the $\chi^2_{\rm min}$ estimator obtained in that fit to the
one of the full fit (here the {\it standard fit}). To assign a probability to the observation, the
$\Delta\chi^2_{\rm min}$ obtained this way must be gauged with toy MC experiments to take into account the
``look-elsewhere'' effect introduced by the explicit selection of the pull outlier. We find that
in $(1.4\pm0.1)\%$ (``$2.5\sigma$'') of the toy experiments, the $\Delta\chi^2_{\rm min}$ found exceeds the
$\Delta\chi^2_{\rm min}=8.0$ observed in current data.

The strong coupling at the $Z$-mass scale is determined by the {\it complete fit} to
$\alpha_S(M_Z^2)=0.1193 \pm 0.0028 \pm 0.0001$ where the first error is 
experimental and the second due to the truncation of the pertubative QCD series. \par

Figure~\ref{sm3} compares the direct measurements of $M_W$ and $m_t$, 
shown by the shaded/green $1\sigma$ bands, with the 68\%,  95\%, and 99\%~CL obtained 
for three sets of fits. The largest/blue (narrower/yellow) allowed regions are derived 
from the {\it standard fit} ({\it complete fit}) excluding the measured values in the 
fits. The inclusion of the LEP and Tevatron Higgs searches significantly impacts the 
constraints obtained. Figure~\ref{sm3} allows to compare the indirect and direct determination 
of the $M_W$ and $m_t$. So far the indirect determinations and the direct measurements are 
in good agreement. The third set of fits (narrowest/green) results from the {\it complete fit} 
including the measured values. Hence, it uses all available information and leads to 
the narrowest allowed region. \par

% oblique parameters
\section{Oblique Parameters}
A common approach to constrain physics beyond the SM using the global electroweak fit is the introduction 
of oblique parameters, which assumes that the contributions of new physics models only appear 
through vacuum polarization. Most of the effects on electroweak precision observables can be parametrized by three 
gauge self-energy parameters ($S$, $T$, $U$) introduced by Peskin and Takeuchi~\cite{peskin}. $S$ ($S+U$) describes 
new physics contributions to neutral (charged) current processes at different energy scales, while
$T$ measures the difference between the new physics contributions of neutral and charged current 
processes at low energies ({\it i.e.}, $T$ is sensitive to isospin violation). 
Further generalizations like additional corrections to $Zbb$ couplings~\cite{burgess} 
can be also taken into account.\par

\begin{figure}
\centering
\subfigure[Contours of 68\%, 95\%, and 99\%~CL obtained from scans of fits with fixed
     variable pairs $M_W$ vs. $m_t$ for three sets of fits. 
     The horizontal bands indicate the $1\sigma$ regions of measurements (world averages).]
   {\epsfig{figure=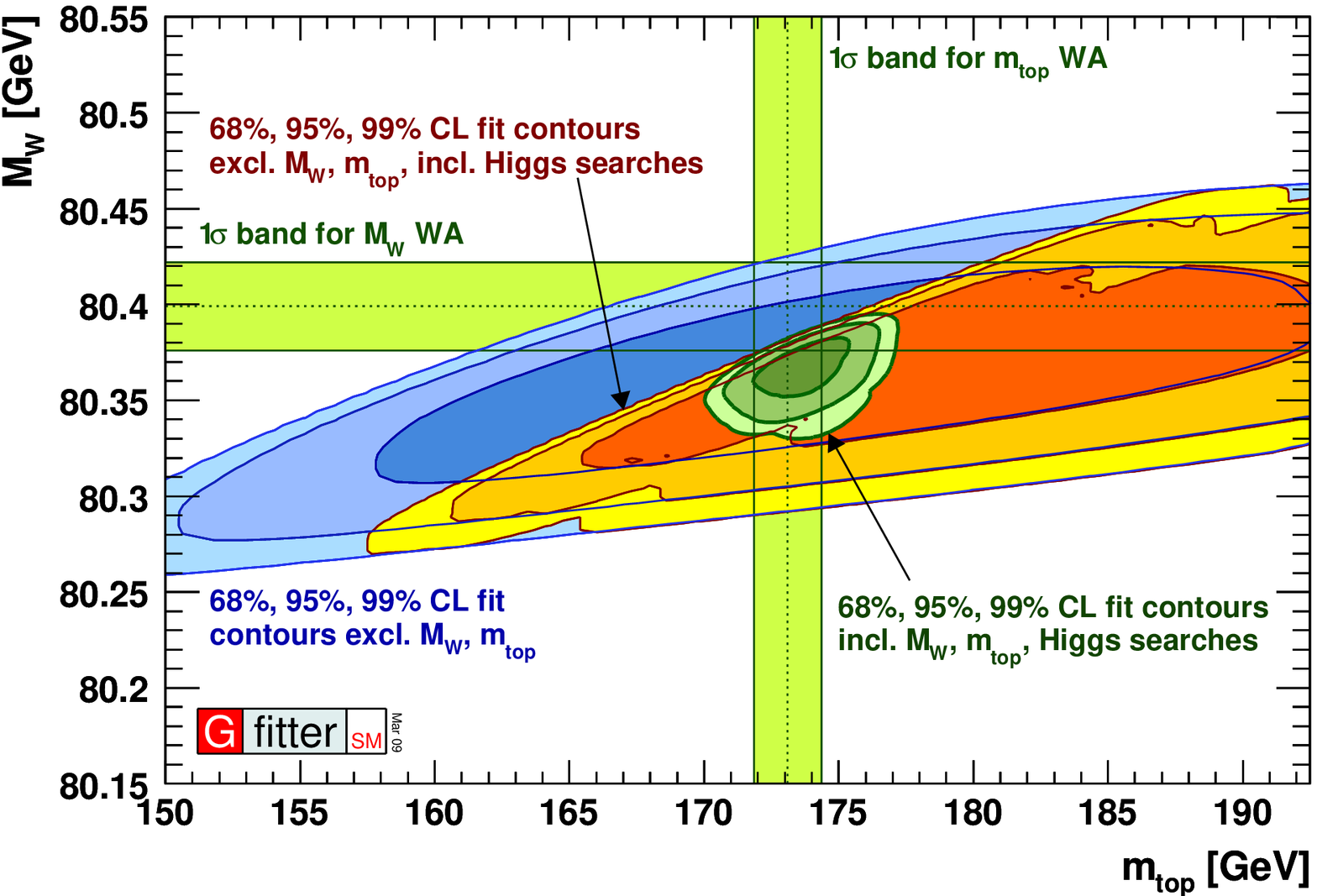,                scale=0.38}\label{sm3}}\hfill
\subfigure[Allowed contours of 68\% and 95\% CL in the ($T$,$S$)-plane obtained from fits with 
      $m_t=172.4$~GeV and $M_H$=116, 350, and 1000 GeV.]
   {\epsfig{figure=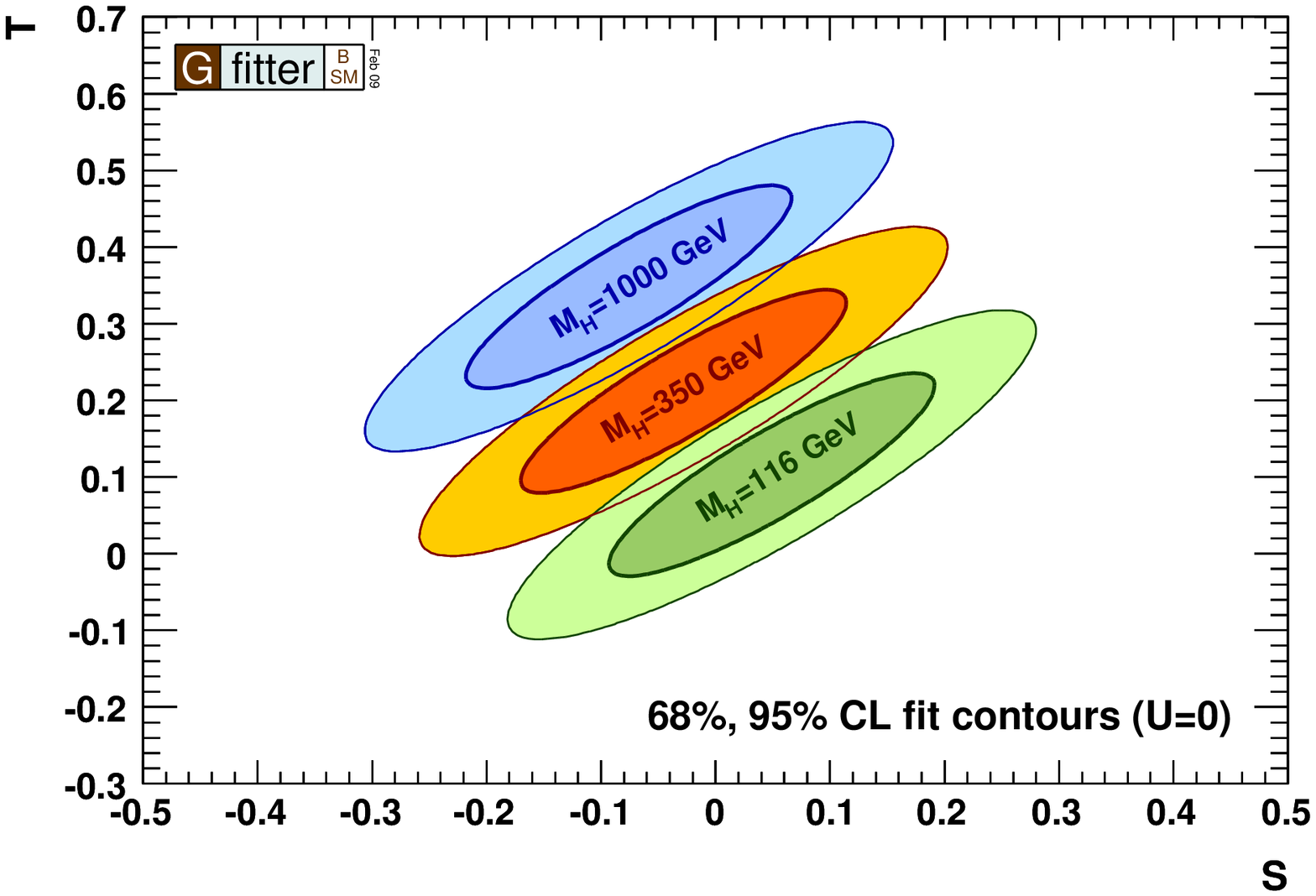,           scale=0.38}\label{stu1}}
\end{figure}

The constraints on the $STU$ parameters are derived from the fit to the electroweak precision data, 
presented in section~\ref{smfit}. The $STU$ parameters replace $M_H$ and $m_t$ as free parameters in the fit. 
The following fit results are determined from a fit 
assuming $\alpha_{S}(M_Z^2)=0.1193\pm0.0028$, $m_t=172.4$~GeV, and $M_H=116$~GeV (in parentheses $M_H=350$~GeV): 
\be
\begin{aligned}
S&=0.02\;(-0.06)\; \pm\; 0.11\\
T&=0.05\;(\ \ \,0.15)\; \pm\; 0.12\\
U&=0.07\;(\ \ \,0.08)\; \pm\; 0.12
\end{aligned}
\ee
Since $U$ is generally small in new physics models and only constraints by the mass and width 
of the $W$ boson, the $STU$ parameter space is often projected to a two-dimensional parameter 
space in which the experimental constraints are easy to visualize. Figure~\ref{stu1} shows
the 68\%,  95\%, and 99\%~CL allowed contours in the ($T$,$S$)-plane for three different assumptions for $M_H$. 
In any case the oblique parameters are small, {\it i.e.}, possible new physics models may affect the 
electroweak observables only weakly.

% Littlest Higgs Model with T-Parity
\section{Littlest Higgs Model with T-Parity}
The fine-tuning problem of the Higgs mass parameter (hierarchy problem) is one of the driving arguments
to consider physics beyond the SM. Besides supersymmetric extensions of the SM, Little Higgs theories 
provide a way to tackle the hierarchy problem. The generic structure of Little Higgs Models is a global 
symmetry broken at a scale $f$ (around 1 TeV) where new gauge bosons, scalars, and fermions exist 
canceling the one-loop quadratic divergences to the Higgs mass from the SM particles. \par

The Littlest Higgs Model (LHM)~\cite{arkani} is based on a non-linear $1\sigma$ model describing an SU(5)/SO(5) 
symmetry breaking. T-parity conservation can provide a possible candidate for a dark matter WIMP (similar to 
R-parity conservation in supersymmetry). 
In addition T-parity forbids tree-level contribution from heavy gauge bosons to the 
electroweak precision observables. The $STU$ parameters of the oblique parameter fit 
are replaced by the calculations of the corresponding ones in the LHM~\cite{hubisz}. The new floating
parameters of the fit are: $f$ the symmetry breaking scale, $s_\lambda \approx m_{T-}/m_{T+}$ in leading order 
the ratio of masses of the T-odd and the T-even state from the LHM top sector, and $\delta_c$ a 
order one-coefficient, which exact value depends on detail of UV physics. The latter parameter is treated 
as theory uncertainty in the fit $\delta_c=-5...5$. \par

\begin{figure}
\centering
\subfigure[Allowed contours of 68\%, 95\%, and 99\% CL obtained from scans of fits with fixed variable pairs 
     $f$ and $M_H$ ($s_\lambda=0.45$). 
     The parameter $F$ is a quantitative measure of fine-tuning 
     (larger values of $F$ correspond to larger degree of fine-tuning).]
   {\epsfig{figure=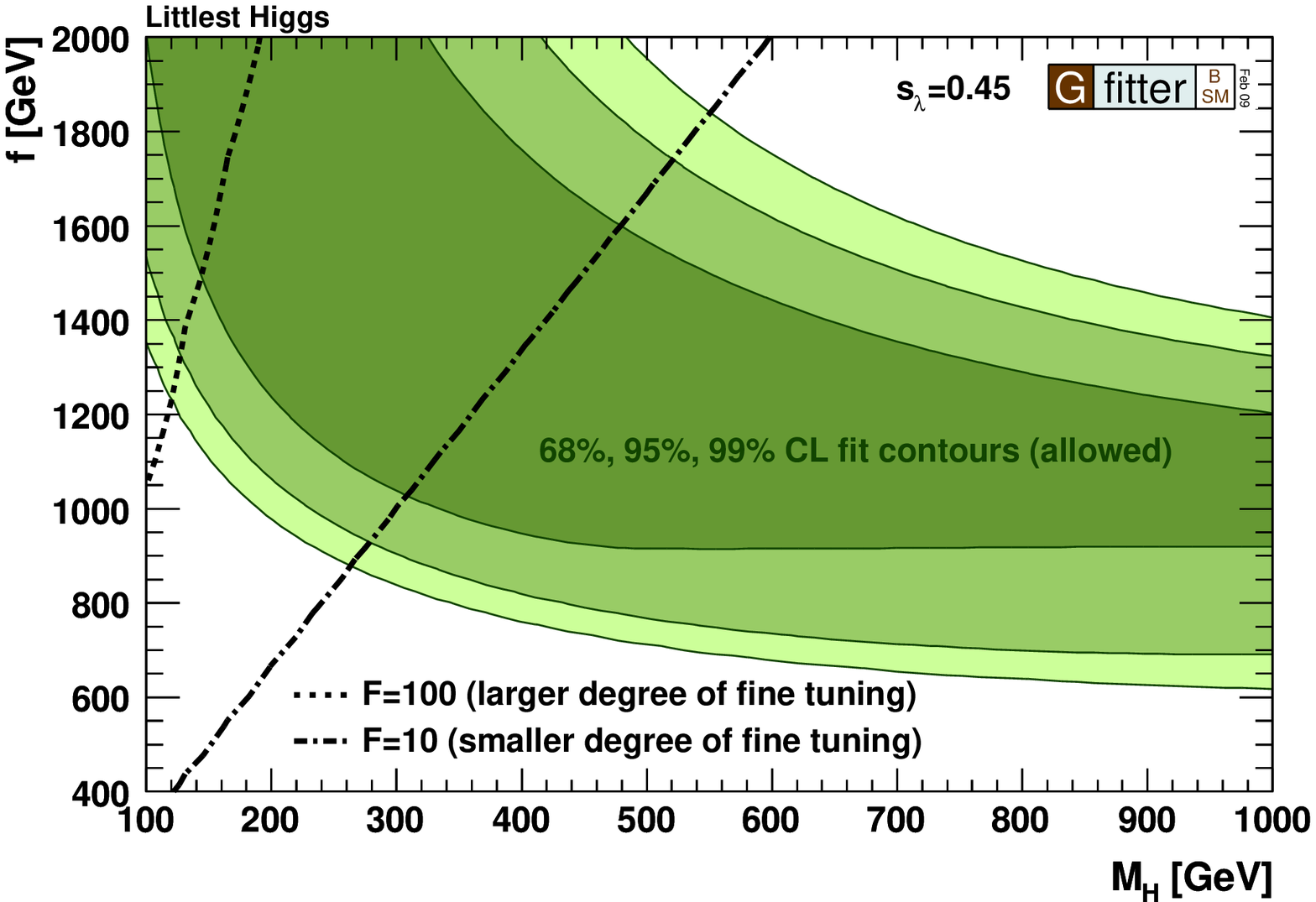,  scale=0.38}\label{stu2}}\hfill
\subfigure[Simple overlay of the 95\% CL exclusion regions in the ($\tan \beta, M_{H^{\pm}}$)-plane from 
         individual 2HDM constraints and the toy-MC-based result (black solid line) from the combined 
         fit overlaid.]
   {\epsfig{figure=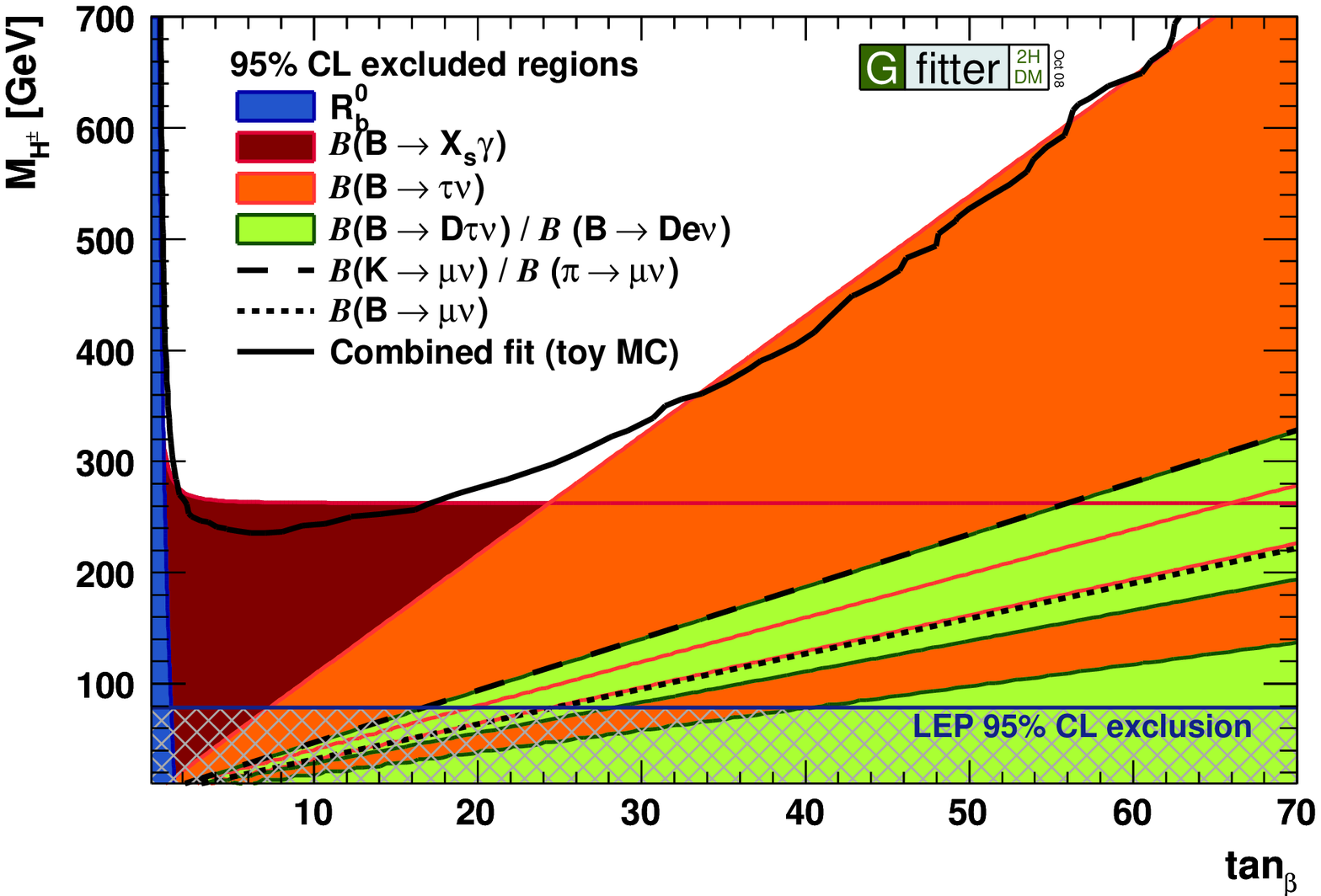, scale=0.38}\label{2hdm}}
\end{figure}

Figure~\ref{stu2} shows the 68\%, 95\%, and 99\% CL allowed contours in the ($M_{H}$,$f$)-plane for a fixed value of 
$s_\lambda = 0.45$. Contributions of the T-odd partners of light fermions to the $T$ parameter are neglected.
This assumption is justified as long as the T-odd fermions are sufficiently light.
The parameter $F$ is a quantitative measure of fine-tuning, indicated by the black lines.
Since larger values of $F$ correspond to larger degree of fine-tuning, large values of $M_H$ are more preferred
than small values. Therefore, large Higgs masses are not only allowed by the electroweak precision data, but they
are also favored in terms of fine-tuning.  

% Two Higgs Doublet Model
\section{Two Higgs Doublet Model}
In the Type-II 2HDM, we constrain the mass of the charged Higgs and the 
ratio of the vacuum expectation values of the two Higgs doublets using current measurements 
of observables from the $B$ and $K$ physics sectors and their most recent theoretical 2HDM 
predictions, namely ${R_b^0}$~\cite{zsummary,haber}, the branching ratio (BR) of 
${B\to X_s\gamma}$~\cite{hfag,misiak}, the BR of leptonic decays of charged pseudoscalar 
mesons ($B\to\tau\nu$~\cite{chang,hou}, $B\to\mu\nu$~\cite{babar,hou} and 
$K\to\mu\nu$~\cite{flavianet}) and the BR of the semileptonic decay 
$B\to D\tau\nu$~\cite{dtaunu,kamenik}. \par

For each observable, individual constraints have been derived in the ($\tan\beta,M_{H^{\pm}}$) 
plane. Figure~\ref{2hdm} displays the resulting 95\% excluded 
regions derived assuming Gaussian behavior of the test statistics, and one degree of 
freedom. The figure shows that $R_b^0$ is mainly sensitive to $\tan \beta$ excluding small 
values. BR($B\to X_s\gamma$) is only sensitive to $\tan \beta$ for values below $\simeq$1. 
For larger values it provides an almost constant exclusion of a charged Higgs lighter than 
$\simeq$$260\,\rm GeV$. For all leptonic observables the 2HDM contribution can be either 
positive or negative since signed terms enter the prediction of the 
BRs resulting in a two-fold ambiguity in the $(\tan\beta,m_{H^\pm})$ space. \par

In addition, we have performed a global fit combining the information
from all observables. 
For the CL calculation in the two-dimensional plane we performed toy MC tests in each scan 
point which allows to avoid the problem of ambiguities in the effective number of 
degrees of freedom. The 95\% CL excluded region obtained are indicated in 
Figure~\ref{2hdm} by the area below the single solid black line. We can exclude a charged Higgs 
mass below $240\,\rm GeV$ independently of $\tan\beta$. This limit increases towards larger $\tan\beta$, 
e.g., $M_{H^{\pm}}<780\;\rm GeV$ are excluded for $\tan\beta=70$.

% Acknowledgments
%\section*{Acknowledgments}

% references

\end{document}